\documentclass{article}

\usepackage{arxiv}

\usepackage[utf8]{inputenc} % allow utf-8 input
\usepackage[T1]{fontenc}    % use 8-bit T1 fonts
\usepackage{hyperref}       % hyperlinks
\usepackage{url}            % simple URL typesetting
\usepackage{booktabs}       % professional-quality tables
\usepackage{amsfonts}       % blackboard math symbols
\usepackage{nicefrac}       % compact symbols for 1/2, etc.
\usepackage{microtype}      % microtypography
\usepackage{lipsum}		% Can be removed after putting your text content
\usepackage{graphicx}
\usepackage{natbib}
\usepackage{doi}
\usepackage{amsmath}

\usepackage{cleveref}
\usepackage[many]{tcolorbox}  

\newtcolorbox{boxA}{
    fontupper = \bf,
    boxrule = 1.5pt,
    colframe = black % frame color
}

\newtcolorbox{boxB}{,
    boxrule = 1.5pt,
    colframe = black % frame color
}

\title{From Incidents to Insights: Patterns of Responsibility following AI Harms}

\author{Isabel Richards \\
	The Leverhulme Centre for the Future of Intelligence\\
	University of Cambridge\\
	Cambridge, UK \\
	\And
        {Claire Benn} \\
	The Leverhulme Centre for the Future of Intelligence\\
	University of Cambridge\\
	Cambridge, UK \\
        \And
        {Miri Zilka} \\
	Department of Engineering\\
	University of Cambridge\\
	Cambridge, UK \\ \\
	\texttt{mz477@cam.ac.uk} \\
}

\renewcommand{\shorttitle}{\textit{arXiv} Template}

\hypersetup{
pdftitle={A template for the arxiv style},
pdfsubject={q-bio.NC, q-bio.QM},
pdfauthor={David S.~Hippocampus, Elias D.~Striatum},
pdfkeywords={First keyword, Second keyword, More},
}

\begin{document}
\maketitle

\begin{abstract}
The AI Incident Database (AIID) was inspired by aviation safety databases, which enable collective learning from failures to prevent future incidents. The database documents hundreds of AI failures, collected manually from the news and media. However, recent criticism highlights that the AIID's reliance on media reporting limits its utility for learning about implementation failures. In this paper, we accept that the AIID falls short in its original mission, but argue that by looking beyond technically-focused learning, the dataset can provide new, highly valuable insights: specifically, opportunities to learn about patterns between developers, deployers, victims, wider society, and law-makers that emerge \textit{after} AI failures. Through a three-tier mixed-methods analysis of 962 incidents and 4,743 related reports from the AIID, we examine patterns across incidents, focusing on cases with subsequent public responses tagged in the database. We identify common 'typical' incidents found in the AIID, from Tesla crashes to deepfake scams.
Focusing on this interplay between relevant parties, we uncover interesting patterns in accountability and social expectations of responsibility. We find that the presence of identifiable responsible parties does not necessarily lead to increased accountability. The likelihood of a response and what it amounts to depends highly on context, including who built the technology, who was harmed, and to what extent. Controversy-rich incidents provide valuable data about societal reactions -- including insights around social expectations regarding responses. Equally informative are cases where expected controversy is notably absent. This work shows that the AIID's value lies not just in preventing technical failures, but in documenting and making visible patterns of harms and of institutional response and social learning around AI incidents. These patterns offer crucial insights for understanding how society adapts to and governs emerging AI technologies.
\end{abstract}

% keywords can be removed
% \keywords{First keyword \and Second keyword \and More}

\section{Introduction}
In an era of unprecedented AI deployment, failure is inevitable -- the key is to learn from it. AI systems, from machine learning classifiers to large language models, are being rapidly deployed across many industries to tackle a range of perceived problems from elderly care \citep{worth2024robots} to traffic optimisation
\citep{carey2023google}. This accelerating adoption is powered by a wealth of investment \citep{mcwilliams2024research} and an international strategic focus \citep{bareis2022talking}. However, the potential for the high-stakes disruption of societies is vast, including impact on education \citep{groza2023brave}, social cohesion \citep{jacobs2024artificial}, and even on concepts such as mourning \citep{lindemann2022ethical}.

The question of \textit{how to learn from failure} is important yet complex. 
%This is not the first time that attempts have been made to learn on a collective level.  
The aviation industry takes an approach of collective learning from failure. Aviation incident databases, such as those hosted by the US National Transportation Safety Board, provide a detailed public record of incidents through extensive investigations carried out by specialist industry teams. Commercial aviation has seen a significant increase in safety over decades, with much of this progress attributed to the use of these shared databases \citep{mcgregor2021}. The aviation industry's approach to improving safety through public databases suggests a straightforward path to improving safety: document incidents so the entire industry can learn from others' mistakes and implement preventive solutions. 

Inspired by the historical success of aviation safety, incident databases for \textit{AI} systems have emerged as community-led efforts to document and learn from failures \citep{mcgregor2021,rodrigues2023artificial}. These efforts share the same vision -- failure must be recorded if it is to be learnt from \citep{mcgregor2021}. The approach also echoes aviation in focusing on the sphere of influence of industry practitioners: to enable developers to learn from others' mistakes and implement preventive solutions. While highly valuable, this vision is complex to realise in practice. AI incident databases, in contrast to aviation, bridge incidents across many industries, and lack the investigatory infrastructure that aviation enjoys. To create a public record of AI incidents, the databases rely on submissions of existing publicly available information about incidents, principally in the form of trade and news media (media reports). Recent criticism \citep{durso2022analyzing,lupo2023risky,shrishak2023deal,turri2023need} highlights that media reports lack sufficient information for developers to avoid the same implementation mistakes.

The AI Incident Database (AIID) – one of the largest of these efforts – has accumulated extensive documentation of AI failures, from patrol robots warding off homeless people (Incident 261, AIID) to chatbot helplines advising those with eating disorders to diet (Incident 545, AIID). In addition to collecting information on AI harm incidents, the AIID began tagging responses to incidents. This is an effort to supplement details on incidents with disclosures from developers and deployers -- a model adopted in cybersecurity. The response, however, has not proven successful to date with very few official responses recorded. 

In this paper, we highlight alternative forms of learning from AI incident databases that avoid these limitations. Instead of focusing on the media reports as a limited source of information about the details of an incident, we approach them as a rich source of information about how different societal actors respond. We applied a three-tier approach to learn from the societal responses to incidents embedded in database entries: First, we quantitatively analysed all 962 incidents and their 4,743 associated reports based on the actors involved – developers, deployers and harmed groups. Next, we used quantitative and qualitative methods to examine the 48 incidents and 163 reports tagged as responses, i.e., with acknowledgement from developers, deployers, or other societal actors. Finally, we combined these perspectives to systematically study sub-categories of incidents, revealing patterns and key factors driving substantive responses -- and lack thereof -- in prominent recurring incidents.

While the statistics of the database are heavily impacted by sampling bias, the methodology developed allows us to meaningfully learn from the incidents that are documented, by comparing similar incidents that did or did not receive responses, and the contextual factors that may have contributed.
We find distinctive patterns in how AI incidents unfold and are addressed. While these patterns are not necessarily representative of all AI incidents due to the limitations of the database, they provide valuable insights into understanding effective societal and legal mechanisms for accountability in varying contexts:

\begin{enumerate}
\item \looseness=-1
\textbf{The identifiability of responsible parties:} Contrary to what might be expected, we show that having identifiable responsible parties does not correlate with better accountability. Rather, many incidents involving major technology companies as both developers and deployers generated non-to-few formal responses. In contrast, incidents with no known developers and deployers, mostly involving deepfakes, show evidence of stimulating the greatest demand for response and legislative action. 

\item \looseness=-1
\textbf{The status of harmed parties:} When an acknowledgement of harms does occur, tentative evidence suggests that in the absence of legislative support,  organisational victims tend to receive more substantive responses (e.g., formal investigation vs. blog post) than individual users, suggesting institutional accountability is shaped more by social and economic factors than technical ones.

\item \looseness=-1
\textbf{The level of public outcry in response to an incident:} We demonstrate that the AIID provides two types of valuable insights about how society negotiates acceptable boundaries for AI deployment: first, from incidents that sparked highly publicised controversies, and second, from incidents that, unexpectedly, gathered no responses. 
\end{enumerate}

Our analysis reveals that the AIID has evolved differently from its original, aviation-inspired, vision. While it may not effectively facilitate direct technical learning for developers and deployers, the database serves an unexpectedly highly valuable resource -- by documenting how different actors respond when things go wrong, we can start to map where socio-legal accountability measures are effective and where there is much left to be desired.

\section{Background and Related Work}

\subsection{AI Incident Databases}
\label{sec:back_AIID}
Since 2018, several initiatives have emerged to document AI-related incidents \citep{atherton2024, rodrigues2023artificial, turri2023need}. These initiatives catalogue cases where AI systems have caused harm or created tangible risk of harm in real-world applications. Among these, two databases have gained prominence and are actively maintained: the AI Incident Database (AIID) and the AI, Algorithmic, and Automation Incidents and Controversies (AIAAIC). Both databases contain hundreds of incidents from publicly available reports, primarily from news and trade media. This format enables public submission of incidents for editorial review while maintaining free data accessibility. Other incident databases efforts include MIT's \textit{AI Risk Repository}\footnote{https://airisk.mit.edu/ai-incident-tracker}, WIRED’s \textit{Artificial Intelligence Database}\footnote{https://www.wired.com/category/artificial-intelligence/}, and Algorithm Watch’s \textit{Tracing the Tracers}\footnote{https://algorithmwatch.org/en/tracing-the-tracers/}.

In this work we focus exclusively on the AIID because the data is most suitably structured for analysis, but is otherwise representative of the core data in both databases.

The AI Incident Database, initiated by Sean McGregor with sponsorship from \textit{Partnership for AI} aims to ``discover and learn from the mistakes of the past" \citep{mcgregor2021}. The database targets corporate product managers, risk officers, and engineers as its primary users, providing consistently catalogued incidents, accessible through a web-based interface \citep{aiid2024}. The aviation industry explicitly inspired the goal of the database: to achieve a significant increase in safety, facilitated by incident reporting (McGregor, 2021). It inherits the focus on avoiding failure in implementation -- inviting an implicit emphasis on technical issues as the cause of failure.

\subsubsection{Learning from implementation failure and its limitations}
Several studies have demonstrated how subsets of incidents from the database can be used to avoid implementation failure in specific technical domains. For instance, researchers have examined failures due to software bugs \citep{kassab2022}, deployed model monitoring \citep{schroder2022}, and threat assessment \citep{tidjon2022}. These studies typically use the database to develop or validate taxonomies for categorising system failures which can be applied to new systems to pre-empt issues. This learning is narrow, however, relating to specific highly technical fields and a small range of available incidents where there is sufficient technical information available.

One 2022 study of an AI incident provides an example of broad and deep learning from publicly available reporting following an AI incident (although it does not specifically use the database as the source of these reports).  The study analyses a range of reports, including media reports, about an autonomous vehicle crash in which the first pedestrian was killed \citep{macrae2022}. \citet{macrae2022} distilled twenty-four detailed sources of risk associated with `structural, organisational, technological, epistemic and cultural’ considerations that can inform future implementations. 
This incident was unusual, however. As a transport accident, it benefited from publicly available investigatory reports from regulatory bodies that uncovered significant implementation details. These reports are absent for most incidents in the database.

Recent literature has critiqued the database's effectiveness in providing implementation failure insights more broadly, arguing that media reports generally lack sufficient technical detail \citep{durso2022analyzing,lupo2023risky, shrishak2023deal,turri2023need}. The AIID team appears to be trying to address this issue: First, through efforts to engage developers and deployers of implicated AI systems to
submit \textit{response reports} which would lend detailed implementation insight \citep{schwartz2022}. Secondly, \citet{pittaras2022} suggests that they are developing a methodology for leveraging expert labelling from the AI safety community to provide speculative intuition into what might have gone wrong technically. However, to date, these have had limited impact: since the 'response' initiative began, only 5.8\% of new incidents have responses, and there are no incidents with expert speculative labelling. This paper focuses on what can be learnt from the AIID as is, without additional technical detail.

\subsubsection{Learning from the database in a way that goes beyond the implementation failure}
In a handful of cases, incident data has been used for learning beyond implementation failure. Scholars have used accumulated incidents to learn about which risks need to be prioritised for investigation, without focussing on detailed technical information. 
\citet{hoffmann2023adding} uses the incidents to understand what information needs to be captured about harm to make cases of AI harm comparable.
Two further studies use manual or machine-learning-based content analysis to establish emergent categories in the incident data: one focusing on AI ethics issues in specific application areas \citep{wei2022ai} and another on which ‘AI tasks’ are vulnerable to which ‘failure modes’ \citep{zhan2023does}. \citet{feffer2023ai} explores the role of AI incident databases in raising awareness of AI harms. Specifically, this study adopted the AIID as an educational tool in a ML undergraduate course and found that it positively influenced students’ understanding of, and sense of urgency around, AI harms.

\subsection{What could be learnt from media reports?}
Media and official reporting of autonomous vehicle crashes have prompted another sort of investigation: sociological studies. A prominent example is \citet{stilgoe2023machine} -- an analysis of a 2016 Tesla Model S crash which resulted in the first recorded fatality involving a self-driving car. The car, in `Autopilot’ mode, failed to see a truck crossing its path and the car’s owner died in the ensuing crash \citep{yampolskiy2016}. This time, the investigation did not focus on technical details, instead, \citet{stilgoe2023machine} interrogates the language of public reporting following the Tesla crash to identify what has been learnt from the incident, and what has been ignored. Observing the different ways actors responded to the incident provides a critical window into \textit{haphazard social learning} -- the way in which ``society and its institutions make sense of novelty”. With new technologies, the work highlights social reactions as an important primer for regulators to decide the distribution of liability and the thresholds of acceptable safety standards. 

While media reports are not always an objective or complete way to understand what happened in the incident, they are an important source of information in revealing attitudes and actions that can be learnt from in their own right. Science and Technology Studies (STS) scholars consider public controversy around the use of sociotechnical systems -- often documented in media --  as \textit{empirical occasions} --researchable events demonstrating relations between a whole variety of actors from science, politics and industry \citep{marres2015mapping}.

\subsection{Responding to failure}
\looseness=-1
Since the beginning of 2023 the AIID team have been explicitly recording one particular type of societal reaction to incidents -- formal public acknowledgement from developers and deployers. 
The importance of response is a key aspect of interpersonal ethics -- understanding that responsibility lies not only in the prevention of failure and wrongdoing, but in acknowledging their inevitability and examining the subsequent patterns of objection, protest, and response \citep{strawson2003freedom,mckenna2012conversation,scanlon2000we,scanlon2008moral}. When we accept that mistakes and wrongdoing are unavoidable parts of human interaction, our focus shifts to how we respond to the harms we cause. This responsibility extends beyond prevention and includes providing appropriate responses when failures occur. The nature of an adequate response \citep{raji2022outsider} is heavily contextual, and depends on the relationship between the parties involved \citep{wertheimer1998constraining}, on who can be held accountable \citep{cohen2006casting,dworkin2000morally}, the degree to which the harm was foreseeable or preventable \citep{coates2016epistemic,rosen2004skepticism}, the involvement of multiple actors in causing the harm \citep{zwart2015responsibility}, and so on. Responses can manifest in various forms, from apologies to compensation of the victims and preventive measures for the future \citep{helm1990richard}.

\section{Methodology}

In this section we describe the dataset, followed by our analytical approach for analysing the data. 

\subsection{The AI Incident Dataset}
The AIID is a repository of reported real-world problems involving AI systems. The formal definition of incident is ``An alleged harm or near harm event to people, property, or the environment where an AI
system is implicated''. Harm is not strictly defined, and contributors are advised to document harm if a ``plausible argument" can be made that a harm occurred.\footnote{The AIID editors provide guidance on recognising harms in the editors guide (https://incidentdatabase.ai/editors-guide/).} 
Each incident in the AIID is captured as a collection of publicly available media reports about the same event. The dataset, as downloaded on the 10th of March 2025, contains 962 incidents and 4743 related reports, submitted since 2019. The number of reports per incident varies significantly, between 1-58 reports per incident (see \Cref{apx:fig_reports_incident}). The dataset consists of two tables, one with basic data about each incident, and another with the corresponding reports and their submission details. The meta-data covers the incident, the implicated parties (victims, developers, deployers), reporting and submission (see \Cref{apx:data}). 

Media reports and the relevant meta-data are collected via a public webform and are reviewed by editors before being incorporated into the dataset. A handful of submitters have submitted hundreds of reports, however, most submit only one (see \Cref{apx:fig_reports_submitter}). The top submitters are, or have been, formally associated with the AIID as editors.

\subsection{Analytical approach}
This research aims to uncover insights within the AIID beyond technical learning by paying close attention to the responses of different societal actors captured within media reports. We employed a three-tier exploratory approach: 1) we analysed all recorded incidents with reference to the actors involved – developers, deployers and harmed groups; 2) We focused on a subset of incidents that received formal acknowledgement from developers and deployers (`responses'); 3) we combined these perspectives to systematically study sub-categories of incidents to reveal insightful patterns in the contextualised responses of different actors. Below are the methodology and limitations of each step: 

\vspace{-0.5em}
\subsubsection{All incidents}
\label{met:def}
To enable quantitative cross-comparison of all incidents with a focus on actors, we used existing data fields for Developers and Deployers and inductively developed categories: 

\paragraph{Developer and deployer}
`Developer' and `Deployer' are data fields in the AIID. The definitions from the database editor guide are:

\textbf{Developer:} the organizations or individuals responsible for producing either the parts or the whole intelligent system implicated in the incident.

\textbf{Deployer:} the organizations or individuals responsible for the intelligent system when it is deployed in the real world.

The ‘Developer(s)’ and ‘Deployer(s)’ fields were used to categorise incidents based on whether the responsible parties were known or not and, if known, whether they were the same organisation/individual; this required limited data-cleaning. For example if  ‘unknown-hacker’ was a recorded deployer, this was categorised as unknown.

\paragraph{Harmed groups}

For each incident, the following information was reviewed: the ‘alleged harmed or nearly harmed groups’ tags (e.g. ,`the-guardian', `family-of-lilie-james']), and the description of the incident. While reviewing, appropriate harmed group categorises were developed inductively. All incidents were then tagged accordingly. Harmed groups are not mutually exclusive. The labelling was done by two authors of the paper and checked for consistency by cross-labelling. \Cref{tab:harmed} describes all of the developed categories.

\begin{table}[h]
\centering
\caption{Harmed group categories}
\begin{tabular}{|p{3cm}|p{4cm}|p{6cm}|}
\hline
\textbf{Harmed Groups} & \textbf{Notes} & \textbf{Examples (where needed)} \\
\hline
User & Harmed as a direct user of AI-enabled software/hardware &  \\
\hline
Participant & Forced to engage in AI-enabled software as a result of engaging with a distinct process & Applying for a job; being a customer in a shop; using social media but being harmed by some other AI incident via the content being shown there \\
\hline
Public & Harmed engaging in ordinary civic processes that cannot be opted out of & Being harmed while walking down the street; engaging in a government application; visiting a hospital; being scammed \\
\hline
Prominent & Harmed as a result of being an individual well known to the public & Being impersonated using deepfakes  \\
\hline
Vulnerable individual & Harmed as a result of protected characteristic & disproportionately effected due to being poor; elderly targeted by scammers  \\
\hline
Employee & Harmed while undertaking a role &  \\
\hline
Organisation & Where an organisation suffers significant harm & Reputational or financial damage \\
\hline
Nature & Explicit harm of nature as opposed to the general harm to the environment implied in all AI use & AI monitoring missed poachers harming rhinos; AI failed to correctly monitor water quality reduction so that rivers could be protected \\
\hline
\end{tabular}
\label{tab:harmed}
\end{table}

\subsubsection{Incident with responses}
\label{meth:responses}
An \textit{AI Incident Response} refers to ``a public official response to an incident in the AI Incident Database from an entity (i.e. company, organization, individual) allegedly responsible for developing or deploying the AI or AI system involved in said incident.'' (Def.1) \citep{schwartz2022}. 
Beginning in 2023, editors have sought tag whether submitted media reports constitute a response. Responses are expected to vary in completeness, but at a minimum, responses involve a proactive and direct acknowledgement of the incident. Responses may also include what happened, why it happened and what is being done to prevent something similar from happening again \citep{schwartz2022}. An incident with a response has at least one report tagged as a \textbf{response}. 
All incidents with responses were analysed qualitatively. This included reviewing and analysing 638 reports associated with 48 incidents. 

\vspace{-0.5em}
\subsubsection{Establishing patterns}
Combining these perspectives, we systematically studied sub-categories of incidents and inductively grouped incidents by common characteristics, identifying `typical incidents' and factors that impact whether a formal response was received.

\vspace{-0.5em}
\subsubsection{Limitations of the dataset}
The incidents captured in the AIID are not a full, nor a representative set of the type and frequency of AI-related harms worldwide. One main reason for this sampling bias is the reliance on media reporting. Incidents published in the media are likely to be those
that are surprising or controversial \citep{dandurand2023freezing,stilgoe2018machine}, and may reflect some harmed groups over others \citep{hilgartner2000science,marres2015map}. Harms that have yet to be well-articulated or investigated \citep{marres2021issues,nixon2011slow} are likely to be excluded. In addition, only a very small fraction of the incidents reported in the media will be submitted to the database, and these are almost exclusively from media reports in English. As such the AIID cannot be viewed as a straightforward `map' of underlying incidents. This is a common challenge
with incident reporting where accumulated data typically reflects reporting behaviour more than occurrence \citep{macrae2016problem}. In recognition of these limitations, the analytical approach above avoids making claims about all incidents. Instead, we focus on learning from the documented incidents by comparing similar incidents that did or did not receive responses to understand relevant contextual factors.

Having acknowledged the aim of learning from a biased subset of incidents, further limitations restrict what can be concluded from the reports themselves. Media reports are not objective accounts of the incidents. They lack detailed information about the implemented systems \citep{turri2023need} and are embedded with assumptions rooted in geography \citep{dandurand2023freezing} and industry \citep{camilleri2023media} (see \Cref{sec:back_AIID}). The number of reports attached to each incident is also not a true measure of the media attention the incident received. While the number of reports submitted per incident was envisaged as a proxy for interest in an incident, in practice this is not the case. Instead reports being submitted by many submitters, it is often the case that a single, prolific submitter, submits many reports for a given incident. The quantity of reports depends on their efforts as well as on the number of available reports. In an informal conversation with one of these submitters, they described how they would stop searching for more reports for an incident when the new articles ceased to provide new or interesting information, or when they ran out of time. Due to these reasons, we choose not to consider the number of reports as a key factor in the analysis, choosing a mixed-method approach to allow a more meaningful comparison between incidents. 

\section{Results}
We start by presenting the quantitative analysis of all incidents in the AIID (\Cref{sec:res_1}), identifying patterns across the database with respect to who is harming and who is harmed. We then focus on incidents with responses (\Cref{sec:res_2}), as these contain richer information on what occurred after the incident. The final section presents the `typical incidents' found in the databases under inductive categories (\Cref{sec:res_3}), giving the reader a concrete sense of the harms and their aftermath. We highlight that the quantitative results presented here are with respect to the AIID, and are not representative of all AI harm incidents. 

\begin{figure}
    \centering
    \includegraphics[width=\linewidth]{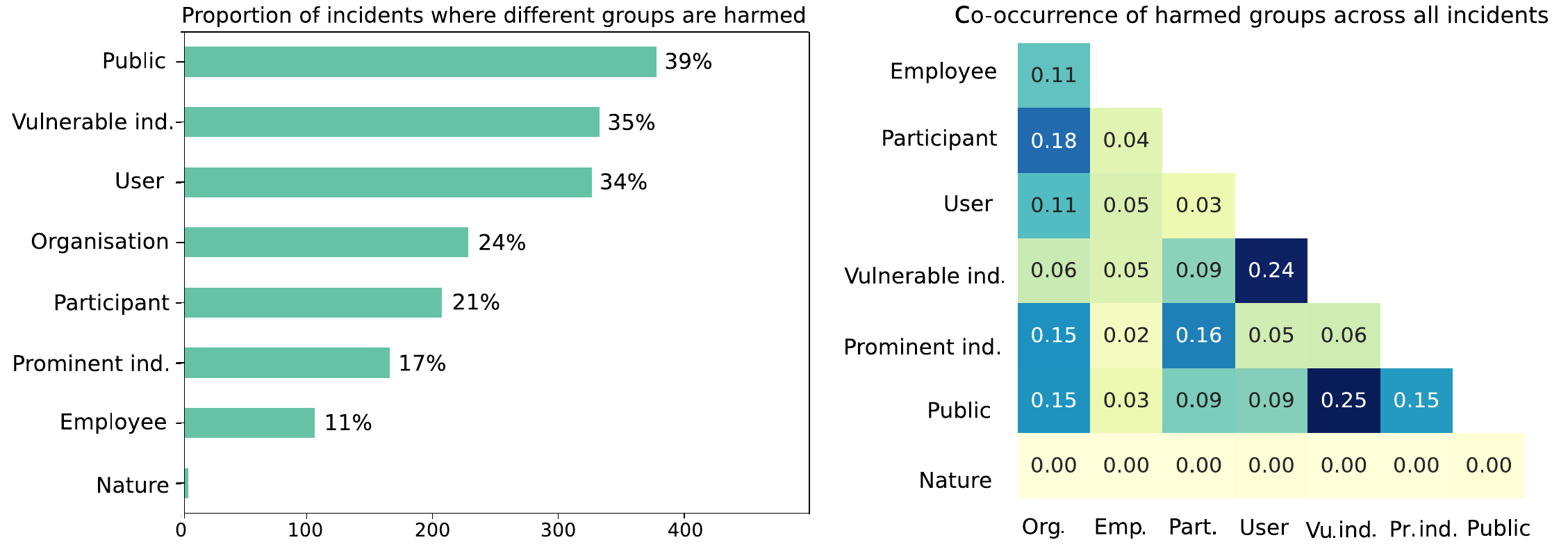}
    \caption{\looseness=-1 \textbf{Left:} proportion of all incidents by harmed group (not mutually exclusive; see \Cref{tab:harmed} for full definitions of all groups). \textbf{Right:} co-occurrence of different harmed groups using the Jaccard Similarity Index (Higher values indicate groups frequently appear
together). }
    \label{fig:harmed_group}
\end{figure}

\subsection{AI harm incidents}
\label{sec:res_1}
We begin by analysing all incidents across two principal dimensions: groups harmed, and
developers and deployers (see \Cref{tab:harmed} for group definitions).

\subsubsection{Which groups were harmed?}
As shown in \Cref{fig:harmed_group} (left), within incidents recorded in the AIID, the public, users and vulnerable individuals were most frequently harmed, with 39\%, 35\% and 34\% of incidents, respectively (see \Cref{tab:harmed} for definitions). The public -- defined as individuals harmed while engaging in normal civic activities for which there is no alternative (e.g., walking down the street, crossing border control, or participating in election, see \Cref{tab:harmed}) -- were harmed the most, in 39\% of incidents. Harmed groups are not mutually exclusive. \Cref{fig:harmed_group} (right) shows the proportion of the different harmed groups and how often they co-occurred in the same incident. We can see that it is not uncommon for vulnerable individuals to be harmed in the same incidents as users or the public. In incidents where the `vulnerable' and `user' tags are both used it is often not only vulnerable users who are harmed. For instance, consider when Yandex, a Russian technology company, released a chatbot which responded to questions with racist and pro-violence replies (Incident 58, AIID). In such cases, harm is not limited to vulnerable users (like marginalised groups) but also impacts non-vulnerable users encountering harmful content, as well as vulnerable individuals who do not use the chatbot but are still impacted by its influence. With incidents where both `vulnerable' and `public' tags are used however, it is common for the person harmed to be in the intersection of these categories, for example where they are harmed as a student: they are classed as `vulnerable’ as a child and as `public’ because they cannot avoid going to school.

\subsubsection{What can be learnt about the developers and deployers?} 

\begin{figure}[ht]
    \centering
    \includegraphics[width=\linewidth]{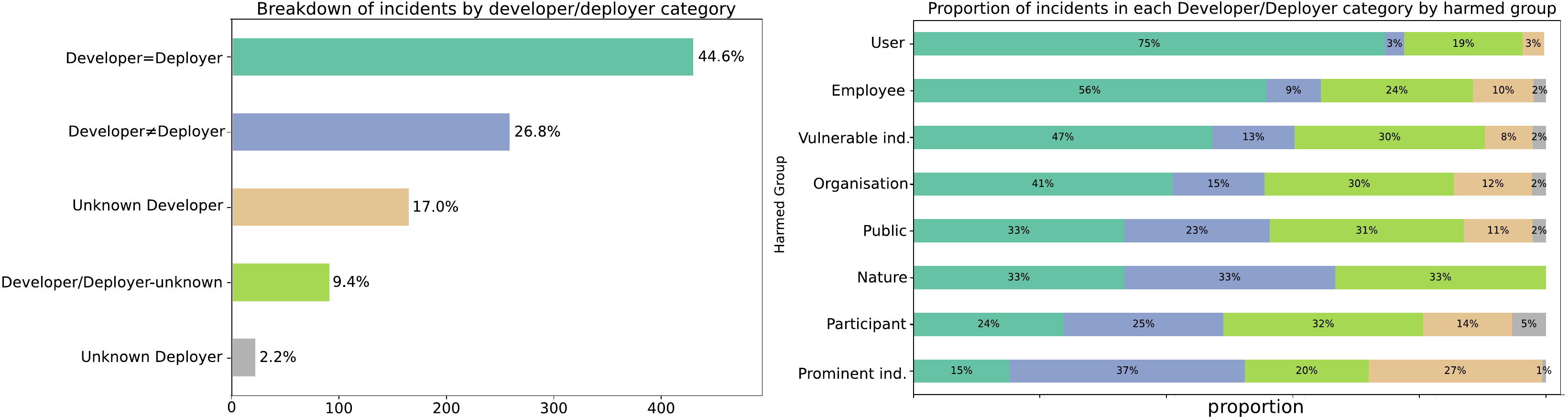}
    \caption{\textbf{Left:} A breakdown of incidents based on whether the developer and the deployer are known and whether they are the same organisation if they are both known. \textbf{Right:} Shown separately for each harmed group.}
    \label{fig:developer_known}
\end{figure}

\Cref{fig:developer_known} (left) shows whether the developer and the deployer are known (see \Cref{met:def} for definitions) and whether they are the same organisation if they are both known. In almost 45\% of incidents the developer is the same as the deployer. Cross-referencing these categories with harmed groups (\Cref{fig:developer_known} right) shows that when users are harmed, it is often  when the developer and the deployer are the same. This naturally invites the question as to whether these organisations are what is commonly referred to as \textit{Big Tech}\footnote{`Big Tech' was categorised manually from a list of 16 prominent global technology companies. The list was developed based on iterative Google searches with terms such as ‘big tech’, and includes: ['alphabet','google','youtube','amazon','apple', 'meta', 'facebook', 'microsoft', 'baidu', 'alibaba', 'tencent', 'openai', 'twitter', 'tesla', 'tiktok', 'bytedance'].}. Indeed, Big Tech companies are quite prominent, with 35\% of incidents having \textit{Big Tech} companies as developers or deployers, and 68\% when the developer and deployer are the same organisation.  

\Cref{fig:developer_known} (right) also raises concerns about third-party developers. In the case where someone is harmed as a \textit{Participant} --  someone who is harmed by an AI system which they did not directly sign up to use -- the developer is not the deployer, or is unknown. We can see that in these cases the developer is often a third-party (25\%) or cannot be identified (32\%). 
An example of this is incident 673 in the AIID, where participants are harmed using Adobe stock imagery because, unbeknownst to them, they are using AI-generated images created by third parties which misrepresent reality.

The category of incidents where neither the developer nor the deployer are known is particularly puzzling (9.4\%). A qualitative analysis of these incidents reveals an unusually high similarity between incidents in this category -- 80\% involve deepfakes. For example, two Canadian residents were scammed by an anonymous caller who used AI voice synthesis to replicate their son's voice asking them for legal fees, disguising himself as his lawyer (Incident 446, AIID).

\subsection{Responses}
\label{sec:res_2}
\looseness=-1
Since the initiative to tag official responses from developers and deployers began, 3.4\% of the submitted reports have been tagged as responses. This corresponds to 163 reports, unevenly distributed among 48 incidents, where one incident has 28 responses, and 25 incidents have a single one (see \Cref{apx:fig_response}). 
Seven of these incidents have five or more responses -- as an official response from the developer or deployer one would naively expect one, or perhaps, two. However, a qualitative analysis of the reports confirmed that 97\% of the tagged responses do not meet the definition of a response (see \Cref{meth:responses}). Instead, the responses fit into three categories: (a) Response of societal actors; (b) Indirect acknowledgement from known developers or deployers, and (c) Official response from developers or deployers, in line with the definition.

\paragraph{Response from societal actors.} Of the incidents with responses, 29\% have no known developer or deployer -- 3 times over-representation compared the their occurrence in the database. 
Instead of the responses of developers and deployers, what was being recorded in these cases was the reaction and activity of different societal actors (e.g., victims, concerned parties, legislators), impacted, but not responsible for, the incident.

Responses of wider societal actors were not confined to incidents with unknown developers and deployers. All incidents with more than one response (23/48) had responses which recorded the reaction of other parties. Based on the definition of responses (see Def.1 \Cref{meth:responses}), this was not originally anticipated by the AIID -- editors set out to tag official responses from developers and deployers but began tagging the responses of wider societal actors. Three incidents (Incidents 597, 616, 645; AIID) had significantly more responses than others. These also had considerably more reports (see \Cref{apx:fig_response}). Upon inspection, this is because these incidents generated public controversy and thus a greater response of societal actors had been captured.

\paragraph{Indirect acknowledgement from developers or deployers.}
Indirect response reports refer to a statement or quote from the responsible party gained via private requests, such as from journalists or regulators, instead of as part of an official statement. These often involved blaming a third party or the user. For example, Sports Illustrated magazine was accused of using AI to generate fake authors and their articles (Incident 616, AIID). The response documents a spokesperson from the magazine’s publisher saying that the particular articles under scrutiny were licensed from a named third party which had ``assured us that all of the articles in question were written and edited by humans'' \citep{Fortinsky2023sports}. Another example of blame shifting occurred when a sepsis prediction algorithm generated significantly higher error rates than advertised when used in a hospital setting (Incident 123, AIID). A developer spokesperson was quoted saying ``their customers… had the ability to tailor the sepsis model to their
specific practice''.

\paragraph{Official responses.} Only 5 out of the 64 reports are official and proactive responses from the developer or deployer. The first incident in this category concerned Microsoft’s chatbot, ‘Tay’, which began tweeting racist, sexist, and anti-semitic comments during the first 24 hours of deployment (Incident 6, AIID). In response, the Corporate Vice President published a blog titled \textit{Learning from Tay’s introduction}. A
second incident involved an ``autonomous security robot colliding with a 16-month-old boy
while patrolling'' a shopping centre (Incident 51, AIID). The response was the organisation's
first ever \textit{Field Incident Report} shared in the trade media.
A more recent incident concerned racially and politically biased outputs from Google's Gemini chatbot. Possibly from a result of ``overcorrection" of biased outputs, Gemini generated racially diverse Nazis, which lead to an statment from Google acknowledging that ``Gemini is offering inaccuracies in some historical image generation depictions", published on X (previously twitter), and saying they are working to fix it. However, the statement did not include an apology per se, just stated that the wide representation was generally good, but here it was ``missing the mark".

In all incidents, there was little scope for deniability, and the responses expressed the intention to learn and iteratively improve. However, they communicated nothing about what was being learnt or what would be done differently, engaging (at least publicly) only superficially in learning from the incident.

\subsection{`Typical' incident categories} 
\label{sec:res_3}
\looseness=-1
The different analytical framings introduced above allowed us to study subsets of the incidents and identify typical incident archetypes and patterns in the circumstances under which they receive responses.

\subsubsection{Developer=deployer `typical’ incident: Big tech incidents where users are harmed.} \looseness=-1 Users were harmed in 76\% of incidents where `big tech' companies were both developers and deployers. 174 incidents fit this category and cover a wide range of harms. Broadly, these can be divided into two subcategories -- incidents relating to social media companies and Tesla incidents.

\paragraph{\textit{Typical incident 1:} Social media incidents.} 
\looseness=-1 These incidents tend to be related to how social media companies present and share information with users. For example, four incidents in this subset relate specifically to TikTok’s \textit{For You} algorithm:

\begin{boxA}
An investigation by NewsGuard into TikTok’s handling of content related to the
Russia-Ukraine war showed its “For You” algorithm pushing new users towards false and
misleading content about the war within less than an hour of signing up. (Incident 185, AIID)
\end{boxA}

\paragraph{\textit{Typical incident 2}: Tesla crashes.} 41 incidents within this category relate to Tesla crashes. In several cases, emergency service vehicles are harmed. For example:

\begin{boxA}
A Tesla on Autopilot mode failed to see a parked fire truck and crashed into its rear on an interstate in Indiana, causing the death of an Arizona woman. (Incident 319, AIID)
\end{boxA}

While the responsible parties are known, harmed users seem relatively powerless in calling for meaningful accountability. Proportionately, there are fewer incidents with responses when the Developer is also the Deployer compared to other categories (see \Cref{apx:fig_response_heat}) -- with only 33\% of the incidents with responses belonging to this category compared to 45\% of the incidents as a whole. Analysing existing responses provides further nuance: There are seven incidents with responses where users are harmed the Developer=Deployer category. In three of these, the organisations were held accountable by legislators, including by invoking existing trading laws (Incident 435, AIID) and data
legislation (Incident 513, AIID). In the remaining cases, the response was mostly insubstantial, from statements of acknowledgement (Incident 642, AIID) to temporarily disabling features (Incident 861, AIID). In the case of the chatbot Tay (Incident 6, AIID; see \Cref{sec:res_2}), where no regulation was applicable, and the only response from Microsoft was a blog post in which it promised to ‘iterate’ on the technology. In contrast, when Microsoft harmed an organisation (Incident 612, AIID), it launched an official probe into the incident within a month.

\subsubsection{Developer$\neq$deployer} 

\paragraph{\textit{Typical incident 3}: Large language models deployed by third-party businesses.} \looseness=-1 `Big tech' developers are much less common when they are not also the deployers, however, this is the case for a significant minority. Typically, these incidents are those where large language models, developed by ‘big tech’, are deployed by third parties such as businesses or individuals. 

\begin{boxA}
The AI-produced, procedural-generated sitcom broadcasted as a Twitch livestream "Nothing, Forever" received a temporary ban for featuring a transphobic and homophobic dialogue segment intended as comedy. (Incident 462, AIID)
\end{boxA}

\paragraph{\textit{Typical incident 4}: Government applications where the developer is known}
This category consists of incidents where the public is harmed and the deployer is a government institution (37\% of all incidents where developer and deployer are known but not the same). One third of these relate to applications of AI systems used by police forces, mostly in similar contexts. For example, 7 incidents relate to police use of \textit{ShotSpotter} -- an AI-based product used to triangulate the location of gunfire, based on sound recording. We note this focus could be due to sampling bias, as many of these reports were submitted by a small team of database editors.

\begin{boxA}
ShotSpotter audios were previously admitted to convict an innocent Black man in a murder case in Chicago, resulting in his nearly-one-year-long arrest before being
dismissed by prosecutors as insufficient evidence. (Incident 255, AIID)
\end{boxA}

Despite clear accountable parties, there were no responses in the cases of LLMs deployed by third parties, and only responses in the case of government applications. This response is attached to two incidents (Incidents 74 \& 529 AIID) delating changes to how Detroit police uses facial recognition following 3 false arrests and one successful lawsuit. The response suggests more safeguarding measurers but double down on the benefits of using the technology. Similarly, most responses obtained in this category relied upon existing accountability mechanisms, including the courts (Incident 608, AIID) and public-sector independent research (Incident 123, AIID). (See \Cref{apx:response_table} for example responses in this category.)

\subsubsection{Both Developer and Deployer are unknown}
Incidents where both the developer and the deployer were unknown make up only 9.4\% of incidents in the database. Despite this small proportion, this category provides valuable insight. 

\paragraph{\textit{Typical incident 5}: Deepfakes shared on social media.} 80\% of incidents in this category involved deepfakes. Of these, the majority shared anonymously on social media. For example:

\begin{boxA}
In Spain, an AI app was used to digitally alter photos of young girls, making them appear naked. This manipulation sparked an investigation after these images were circulated in Almendralejo, a town in the Extremadura region, raising serious concerns about digital privacy violations and the potential spread of these images on pornographic sites. (Incident 610, AIID)
\end{boxA}

The prominence of this type of incident suggests deepfakes shared on social media are particularly liable to generate incidents where there are no identifiable accountable parties. Moreover, social media companies play a role in obscuring developers and deployers in these cases: deepfakes can be shared
without the source being declared and under an unverifiable pseudonym.\footnote{Some social media platforms (e.g., TikTok) have policies requiring “disclosure of synthetic or
manipulated media” \citep{Lovejoy2023TikTok} others (e.g., Meta) have banned them \citep{Straits2023News} although this does not appear to prevent
sharing in practice.} 
The responses in this category captures the reaction of wider societal actors rather than responsible parties.
The responses suggest that the powerlessness created by a lack of recourse contributed to the mobilisation of different societal actors in response to
the incidents. For instance, when Tom Hank’s likeness was used in deepfaked
advertisements (Incident 606, AIID) he is reported to have said: ``[T]here [are] discussions going on in ... to come up with the legal ramifications of my face and my voice --- and everybody else's --- being our
intellectual property''.

Incident 597 (see case study) is amongst the most prominent examples of strong societal response. This incident -- in which nude deepfakes of female students were circulated in a New Jersey high school -- has 21 responses (see \Cref{apx:597}). Importantly, there is sufficient granularity such that you could begin to map out the actions of various actors (victims, communities, the school, the police, the county, legislators) following the incident and the subsequent impact on policy. 
\vspace{-0.2em}

\paragraph{\textit{Typical incident 6}: Scams using Deepfakes} There are 44 incidents in this category reporting on attempted or successful scams, and in almost all of these, scammers used deepfakes. Many of the deepfakes are of prominent individuals, including romantic scams, in which the victim believed they were in a romantic relationship with a famous individual. For example: 

\begin{boxA}
Scammers allegedly used AI-generated image manipulation tools, along with fake social media and WhatsApp accounts, to reportedly impersonate actor Brad Pitt and convince a French interior decorator that she was in a romantic relationship with him. Over 18 months, they allegedly fabricated selfies and messages, reportedly leading Anne to divorce her husband and transfer \$850,000 under the false pretence that Pitt needed money for kidney treatment while his accounts were frozen. (Incident 901, AIID)
\end{boxA}

\begin{figure}[bh]
\begin{tcolorbox}[colback=blue!5!white,colframe=blue!75!black,title=\textbf{Case Study: AI-Generated Fake Nudes Circulated at New Jersey High School (Incident 597,AIID)}]

This incident involves the image-based sexual abuse of female students from a high school in Westfield, New Jersey. AI-generated fake nude images of those students were circulated amongst the student body. The incident, occurring during the summer and brought to the school’s attention on October 2023, has sparked widespread concern among parents, students, educators, lawmakers, and the public. Despite a lack of concrete evidence of the images’ existence, the incident triggered a police investigation, calls for legislative action, and a broader societal debate about the dangers of generative AI misuse and the gaps in legal recourse for victims. \\

After the incident came to light, the first reported response was a letter from one of the parents to other parents calling them to connect and demand a better response. The second is a report, written by a group of parents, of the interaction between the parents and the school following the incident. Parents organized meetings and voiced demands for preventative measures, such as AI misuse awareness training for students. Despite the incident occurring outside school and during the summer, beyond the school’s direct jurisdiction, it was expected that the school will respond strongly to the incident. The school acknowledged the incident and took some steps to address it. The school’s actions included addressing the deletion of the alleged images and educating students about the consequences of AI misuse. The School Principal's ``pledge to raise awareness marks the beginning of what could be a widespread educational effort to prevent such abuses of technology.''\\

The police launched an investigation into the incident, but their efforts were hampered by the absence of concrete evidence, as no images had been recovered. Reports indicated that some students and staff had seen the images in private group chats, but the lack of access to the images complicated efforts to trace their origins or identify the perpetrators.  The local council also publicly acknowledged the incident, which marked the beginning of a broader public discussion. Notably, reports from the UK started to pick up the story after the council’s acknowledgment, suggesting a growing international interest in the issue.\\

The media played a critical role in amplifying the incident and framing the public debate. Articles ranged from basic reporting on the police investigation to in-depth op-eds exploring the implications of generative AI misuse. Opinion pieces called for stronger legal protections and societal awareness about the dangers of deepfakes ``State lawmakers should be making the creation and distribution of fake pornography illegal''. \\

A Westfield High School student, 14, who said she was among more than 30 female students whose photos were manipulated and possibly shared publicly, and her mother, have expressed frustration over what they say is a ``lack of legal recourse in place to protect victims of AI-generated pornography''. The student launched a website and charity focused on supporting victims of AI misuse and raising awareness about its risks. She and her mother also met with lawmakers and participated in advocacy efforts in Washington, D.C., helping to push for new legislation. While the U.S. Justice Department claims this kind of content would be prosecutable under existing federal child pornography laws that cover drawings and cartoons depicting minors engaged in explicit sex, it can’t point to a single prosecution for AI child porn under this legislation.
Following the incident, New Jersey State committed to drafting new laws to criminalize the creation and sharing of AI-generated fake nudes. Lawmakers recognized the incident as a catalyst for accelerating conversations about regulating AI and protecting victims. At the federal level, U.S. Representatives Joe Morelle (D-NY) and Tom Kean Jr. (R-NJ) reintroduced the ``Preventing Deepfakes of Intimate Images Act,'' which would criminalize the non-consensual sharing of AI-generated intimate images and require AI tools to include clear disclosures indicating generated content. The bipartisan nature of this effort underscored the urgency of addressing the issue.
\end{tcolorbox}
\end{figure}

\section{Discussion}
The original goal of the database, inspired by aviation, was to use it to avoid recurring implementation mistakes. Encouraging developers and deployers to submit `official responses’ is one of two ways the AIID is trying to augment the database with implementation details to enable technical collective learning. However, in contrast to the aviation case -- in which investigatory details are provided by dedicated industry teams -- the success of this endeavour relies on the cooperation and commitment of the industry to collective learning. So far, our findings suggest that the approach of seeking proactive, substantive responses from developers and deployers is not fruitful: None of the responses submitted appeared to be submitted directly, as hoped, by developers or deployers; and even the few responses which met the core definition (5/163) lacked the substantive technical detail required for effective learning. 

We claim that currently, the primary value of the database is not learning to avoid implementation failure, but rather learning about the state of both incidents and the responses from different actors in the wake of AI harm. 
Our results demonstrate the value of the AIID as a means of documenting how accountable and non-accountable parties are responding to incidents, and how wider societal actors call for action. 
We take inspiration from interpersonal ethics in focusing not on the avoidance of wrongdoing, but on its inevitability. 
Once we accept that it is impossible to live a life completely devoid of mistakes and wronging others, what becomes central to responsibility is our responsiveness to the resulting harms we do \citep{strawson2003freedom,mckenna2012conversation,scanlon2000we,scanlon2008moral}. All responsible actors are subject to requirements not only to avoid wronging others where possible but to provide an adequate response when they inevitably fail to do so. 

What constitutes an adequate response, and whether it is achieved depends on contextual factors: the nature of the relationship between the wrongdoer and wronged party\citep{wertheimer1998constraining}; standing to blame and hold accountable \citep{cohen2006casting,dworkin2000morally}; the avoidability and foreseeability of the harm and additional failures of responsibility that lead to the wrong taking place \citep{coates2016epistemic,rosen2004skepticism}; the role of other actors in causing the harm \citep{zwart2015responsibility} and so on. The response itself can take many forms: contrition, apologies, recompense, and attempts to avoid future incidents, amongst others \citep{helm1990richard}.

The database proves as a vital record of not just \textit{who} is accountable but also successful and unsuccessful attempts in the practice of \textit{holding} them to account. Our approach demonstrates the database as a valuable resource to interrogate these responses and the lack thereof, reflecting on accountability in practice. This answer calls from the algorithmic accountability literature: worries that the opacity of algorithms \citep{burrell2016machine} and increasingly complex barriers to accountability \citep{cooper2022accountability} are “provid[ing] an easy excuse for irresponsibility” \citep{stilgoe2023machine}. Our work adds to scholarship seeking to understand the practical limitations of accountability in different circumstances by scrutinising and studying real-word cases \citep{stilgoe2023machine,barocas2013governing,raji2022outsider} as a means to open up practical, social or legal, ways forward \citep{ananny2018seeing}.

Our findings reveal patterns not only in the responses of developers and deployers, but also in the response of wider societal actors in demanding an adequate response. These patterns highlight factors that appear to impact the likelihood of a substantive response in prominent incidents.

\paragraph{The existence of an known accountable party does not increase the likelihood of a response to an incident. In fact, there is some evidence to the contrary.} In 71.5\% of incidents in the database the developer and the deployer are known. Intuitively, and perhaps naively, these are incidents where we would expect a substantive response to be more likely -- the responsible organisations or individuals can, in principle, provide more context, technical detail and an indication of what can be done to avoid the issue in the future. The results show that practice this is not the case in: the Developer=Deployer category has a large proportion of all incidents but a smaller proportion of those with responses (45\% vs. 33\%, respectively). It is the opposite when both the Developer/Deployer are unknown -- despite there being no known accountable party, and perhaps even because of it, the demand for a response from wider societal actors is more significant.

Even when an accountable party responds it does not always reflect accountability. For example, Microsoft’s response to the Tay incident (Incident 6, AIID; see \Cref{sec:res_2}) illustrates how responses can be used opportunistically, by embedding implicit opinions about the terms of trial-and-error that the public ought to accept \citep{marres2021issues}. Explaining how the product was tested in the Tay case, Lee says: ``once we got comfortable with how Tay was interacting with users [through user studies], we wanted to invite a broader group of people to engage with [it]. It’s through increased interaction where we expected to learn more…the logical place for us to engage with a massive group of users was Twitter”. The widely accepted view that it is important to learn incrementally through seeking a broader group of users, is used to smuggle in the much more contentious view that a ``massive'' group of users is the appropriate next step. By claiming this step is logical, the implication is that it ought to be accepted. Similar patterns of language have been found elsewhere, with organisations framing new experimentation as “mere baby steps” \citep{borup2006sociology,stilgoe2023machine}. These are ways in which the narrative following incidents is co-opted by implicated parties as an opportunity to prepare society for the innovation they intend to carry out \citep{marres2021issues}, rather than reflecting genuine accountability.

\paragraph{The extent of response from accountable parties depends on who is harmed and whether deployment is direct or third-party}

Consider the case in which Big Tech is both developer and deployer and users are harmed. When the harms impacting users were under the jurisdiction of a regulator, and top-down pressure was applied, Big Tech companies were forced to engage with their roles as accountable actors (see \Cref{sec:res_2}). When the harms impacting users were not covered by regulation, engagement was either not forthcoming or it was superficial (see \Cref{sec:res_2}). This highlights how power can impact accountability: circumstances can arise in which accountable actors are sufficiently powerful such that merely exposing a shameful incident is not enough to compel them to a substantive response \citep{ananny2018seeing}. Explicit regulator pressure does not seem to be required, however, when the harmed party were organisations. For example, when Microsoft harmed users it produced a blog post, but when it harmed an organisation it launched an internal inquiry (see \Cref{sec:res_2}). This suggests that when incidents are exposed, organisations might be in a better position of power, which might compel a more substantive response from Big Tech than users.

The Developer$\neq$Deployer category reveals further patterns in the lack of responses. \textit{Typical incident 3} involved large language models being deployed by third-party businesses. Despite its prominence, no responses from Big Tech were found. In contrast, there was several responses recorded when the incident was similar but Big Tech was both developer and deployer of the LLM (Incident 6 \& 645, AIID; see \Cref{sec:res_2}). This suggests that the shift to third party deployment of LLMs might have decreased the degree to which Big Tech are held accountable as deployers. There are examples in which third-party deployers have been forced to respond substantively. These cases involved vulnerable individuals and existing bottom-up accountability mechanisms such as via the courts (see \Cref{sec:res_3}). These mechanisms are unlikely to be useful for holding Big Tech accountable in response to the growing set of harms caused by third party deployment of generative AI \citep{incidentdatabaseChatGPTIncidents}, where harmed users are often neither vulnerable nor protected in the relevant ways by regulation. In Europe, the EU AI Act specifically designates responsibilities to developers and deployers. While it is too soon to measure its impact on the actions of different actors, this could be an area of valuable future study using the database.

\paragraph{The data is rich where there is controversy, but the lack of controversy where we might expect one is also a valuable source of learning.}

Counterintuitively, incidents where there is no known developer and deployer provide the richest insight into contexts with a substantive response following AI harm. This richness stems from the public controversy stimulated in a number of these cases. As shown in the case study (see blue box), when deepfakes were anonymously shared on social media, the incident reports captured significant response from a wide variety of societal actors -- from victims to local organisations to legislators. The demands, actions, and promises of different societal actors provide a window onto a process of \textit{haphazard social learning} whereby society works out the ``terms'' of responsibility with regards to new technology \citep{stilgoe2023machine}. While doubtless fuelled by an awareness of deepfakes as a new and severe threat \citep{deruiter2021distinct}, it was the lack of recourse that seemed to motivate the societal response, and shape the legislative demands that result.

The debate around deepfake incidents did not identify the role that social media companies play in obscuring the developers and deployers responsible. Public discussion centred on the lack of recourse, with the legislative effort focused almost exclusively on holding the perpetrator responsible. However, social media companies play a critical role in obscuring perpetrators. Our analysis revealed how sharing practices obscured responsibility in most incidents involving deepfakes shared on social media. This contributes to the `many hands’ discussion within the algorithmic accountability literature -- the common use of open-source libraries and pre-trained models complicates the questions of responsibility \citep{nissenbaum1996accountability,cooper2022accountability}. Here we identify an additional problem: how some `hands' can obscure others. In this case, the private mode of sharing has obscured the developer, deployer and the human perpetuator. This highlights that in the case of AI incidents, the question of who to demand a response from, and in what ways, is nuanced, and depends on the specific context and technology involved.

A shift in controversy around similar incidents is also enlightening. \textit{Typical incident 2} (\Cref{sec:res_3}) are Tesla crashes in which a Big Tech company were implicated. In the literature, the response of societal actors to autonomous vehicle incidents has been criticised -- a response is demanded from the wrong actor --  the human operator is typically blamed, despite little control over the behaviour of the system \citep{elish2019moral}. However, we see an interesting pattern where a prominent sub-group of autonomous vehicle incidents caused a different level of controversy and a call for accountability. These were incidents involving emergency vehicles. For example, when a Tesla on autopilot crashed into a parked police car in California, journalists blamed Tesla and not the driver:``This marks the third time a Tesla on Autopilot has been reported to have hit a parked emergency vehicle…The real question is why Tesla is allowed to provide a system it admits is a beta and not a fully-tested product.” (Incident 323, AIID).

Categories where there is an apparent lack of debate are also illuminating. For instance, worryingly, there was only one response found in \textit{Typical incident 4}, where the public was harmed by governments, although harms were substantial (e.g., wrongful arrests). An analysis of the words and actions within these reports could be performed to understand why such incidents have not stimulated greater debate. A similar approach was used to discover that a public challenge relating to the implementation of AI had ceased in Canada due to a convergence of expectations around the technology \citet{dandurand2023freezing}. Noticing a lack of debate is important: allowing harm to go unchallenged and risks them becoming ‘just another fact of life’ \citet{stilgoe2020driving}.

\section{Conclusion}

In this paper, we demonstrate the AIID's potential as a resource to understand where and by whom harms occur; where substantive responses were achieved and through which mechanisms; what responses look like in particular circumstances, and when they are missing. The patterns of response (or lack thereof) we unveil provide insights about how society negotiates responsibility for AI harms. Our contribution is two-fold. Firstly, we establish a revised approach to using the AIID as a rich resource for pragmatic socio-technical learning about effective accountability mechanisms. This approach avoids existing criticisms: while media reports lack detailed implementation information, they contain insights into how actors respond and when accountability is achieved in practice. 
The second contribution are the recurring patterns in the factors affecting the likelihood of achieving a substantive response from accountable actors following AI harm.
We believe this approach will be valuable for scholars studying algorithmic accountability, technology governance, and institutional responses to technological harm. For policy makers and regulators, the findings highlight important gaps in current accountability frameworks, and how such datasets can be used to monitor how these evolve with new regulatory measures. 

\subsection{acknowledgments}
MZ acknowledges support from the Leverhulme Trust grant ECF-2021-429.

\bibliographystyle{unsrtnat}
\bibliography{Arxiv/bib}  %%% Uncomment this line and comment out the ``thebibliography'' section below to use the external .bib file (using bibtex) .

\newpage
\appendix
\section{Key data fields for each table in the AIID}
\label{apx:data}

\textbf{Reports:}
\begin{itemize}
    \item Authors of the report
    \item Description and text of the (news) report
    \item Date published
    \item Report URL
    \item Source domain
\end{itemize}

\textbf{Report submitters:}
\begin{itemize}
    \item Date submitted
    \item Submitter (can be anonymous)
\end{itemize}

\textbf{Incidents:} 

\begin{itemize}
    \item Date (of incident)
    \item Description of the incident
    \item Alleged developer(s) (of system)
    \item Alleged deployer(s) (of system)
    \item Alleged harmed or nearly harmed parties (organic tags, i.e., the submitter based it on reading of the articles. They do not meet a predefined taxonomy.)
    \item Reports relating to incident (list)
\end{itemize}

\newpage
\section{Reports per incident}
\label{apx:fig_reports_incident}
\begin{figure}[h]
    \centering
    \includegraphics[width=0.8\linewidth]{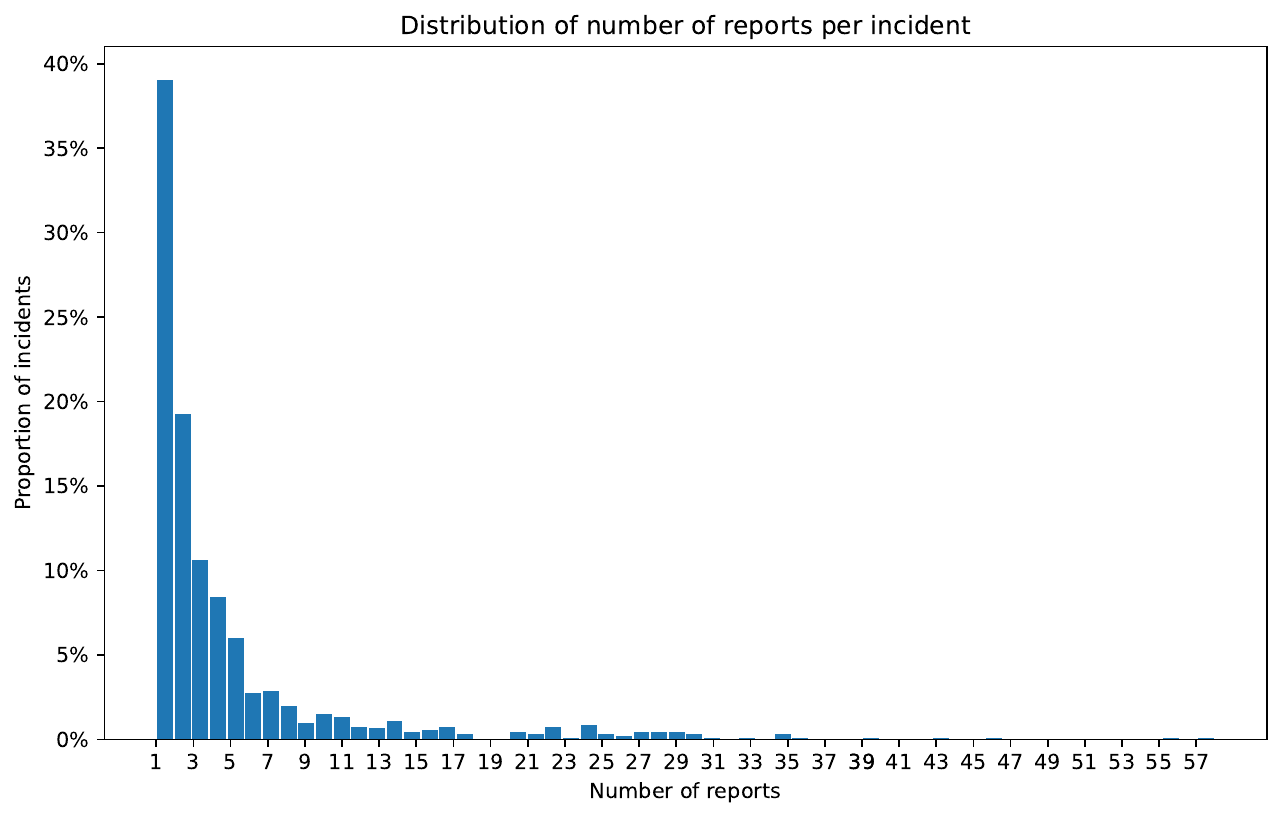}
    \caption{A distribution of number of reports per incident}
\end{figure}

\newpage
\section{Reports by submitter}
\label{apx:fig_reports_submitter}

\begin{figure}[h]
    \centering
    \includegraphics[width=0.7\linewidth]{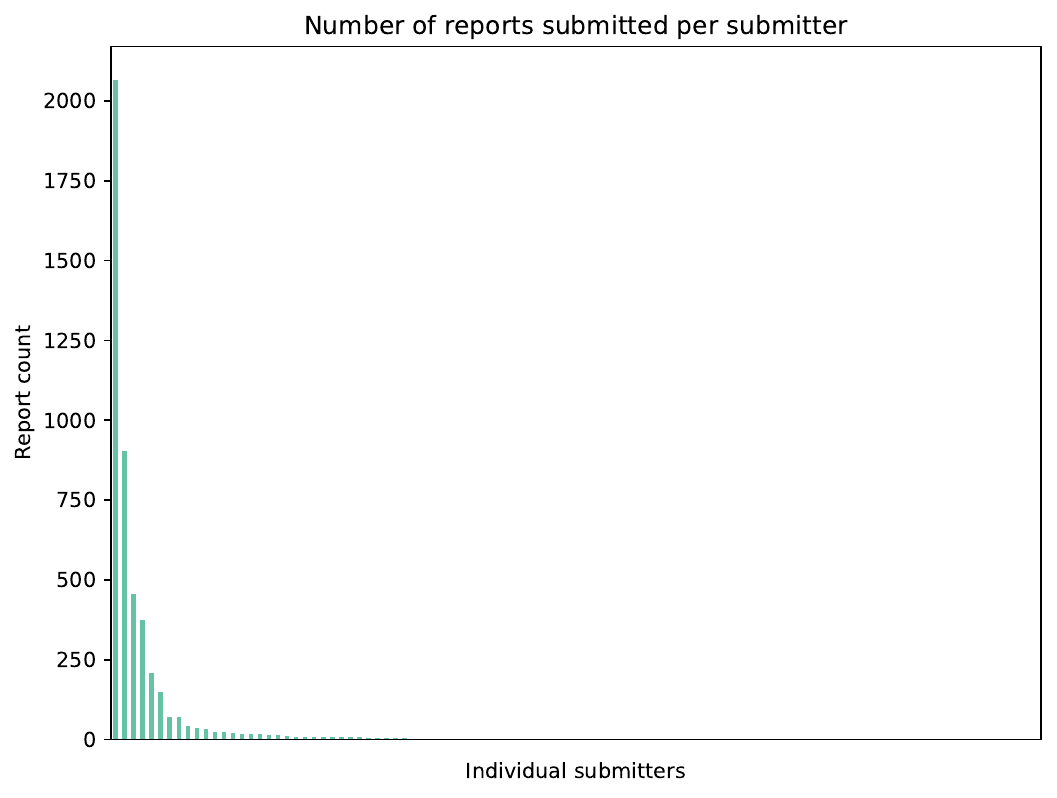}
    \caption{The number of reports submitted by individual submitters}
\end{figure}

\begin{figure}[h]
    \centering
    \includegraphics[width=0.7\linewidth]{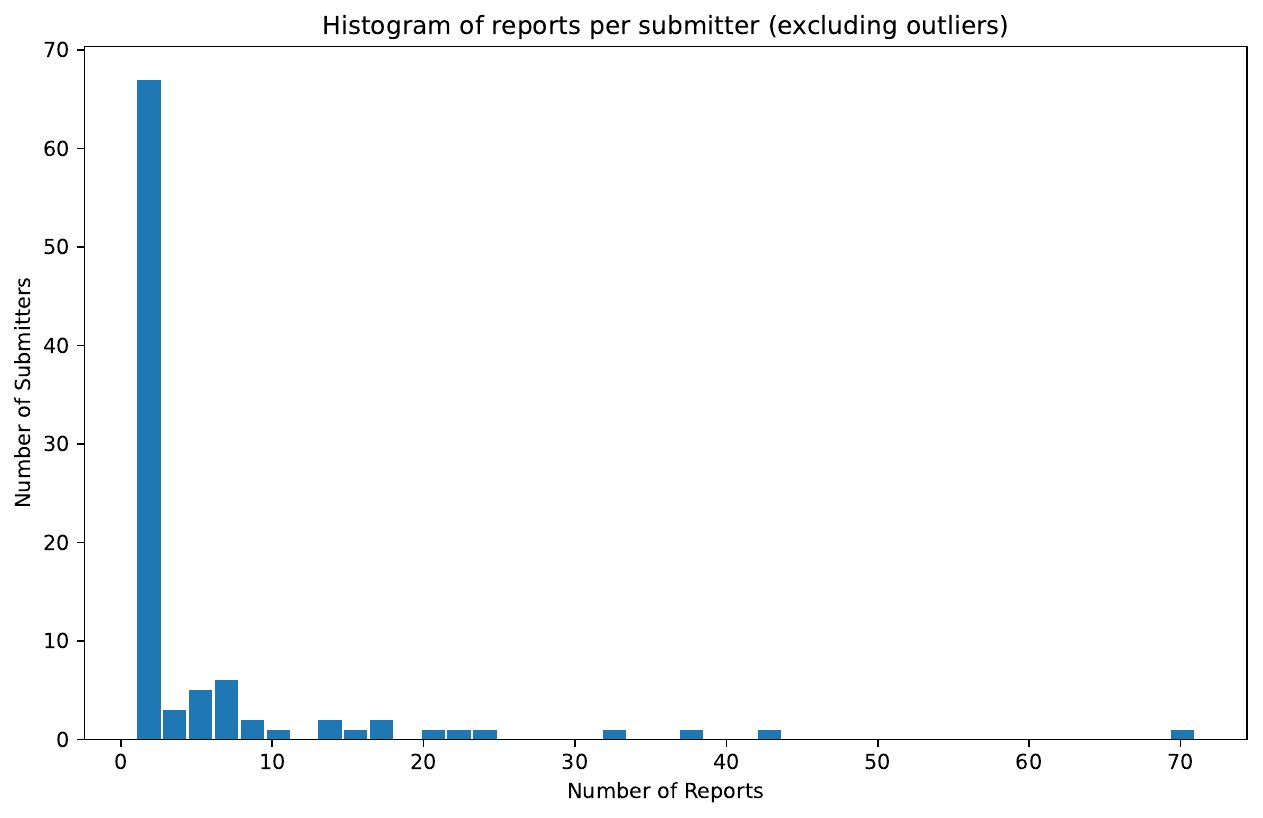}
\caption{A histogram of reports per submitter. Outliers are excluded.}
\end{figure}

\newpage
\section{Incidents with Responses}
\label{apx:fig_response}

% \begin{figure}[h]
%     \centering
%     \includegraphics[width=\linewidth]{Apendix_figs/Incidents_with_responses_Number_of_response_reports.pdf}
%     \caption{The number of responses reports for each incidents with at least one response.}
%     \label{fig:enter-label}
% \end{figure}

\begin{figure}[h]
    \centering
    \includegraphics[width=\linewidth]{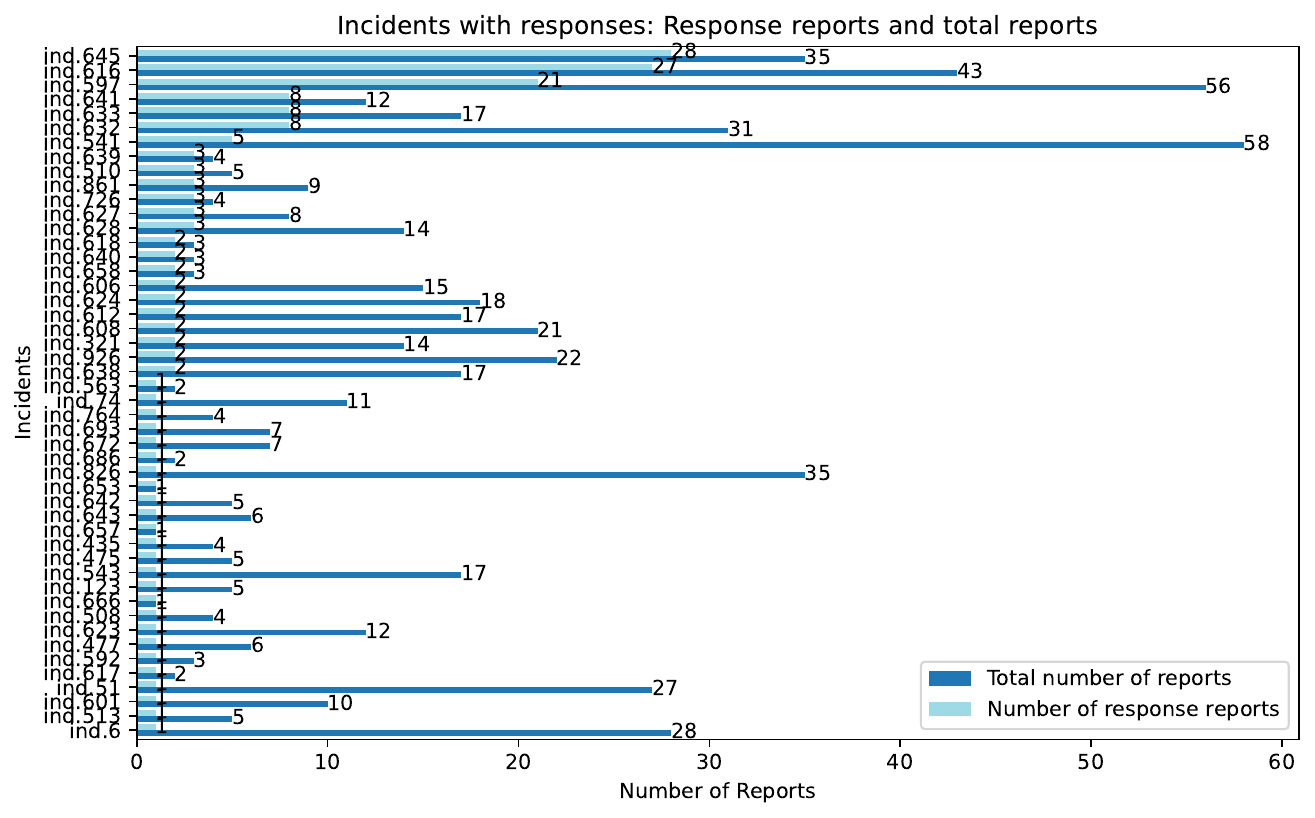}
    \caption{The number of reports and responses reports for each incidents with at least one response.}
\end{figure}

\newpage
\section{Responses by Developer/Deployer category}
\label{apx:fig_response_heat}

\begin{figure}[h]
    \centering
    \includegraphics[width=\linewidth]{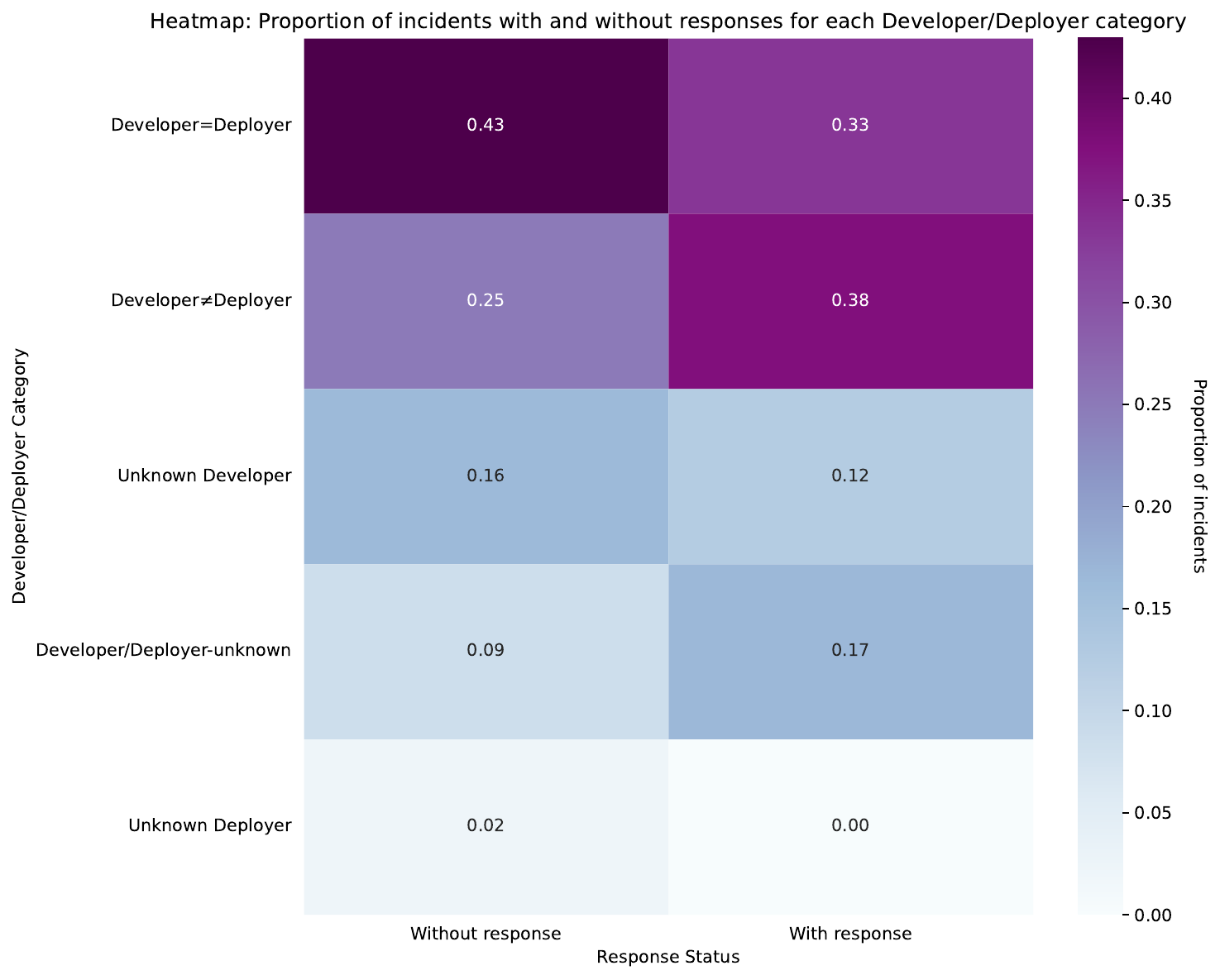}
    \caption{Proportion of incidents with and without responses by Developer/Deployer category.}
\end{figure}

\newpage
\section{Responses when the developer and deployer are known but not the same}
\label{apx:response_table}
\begin{table}[h]
    \centering
    \begin{tabular}{|p{7cm}|p{7cm}|}
\toprule
Title & Notes \\
\hline
Security Robot Rolls Over Child in Mall &
Undeniable physical harm to child \\
\hline
*Epic Systems’s Sepsis Prediction Algorithms Revealed to Have High Error
Rates on Seriously Ill Patients &
Revealed by researchers based at a hospital \\
\hline
Celebrities' Deepfake Voices Abused with Malicious Intent &
Responsible start-up tweets with ideas to 'iterate' safeguarding features \\
\hline
Bing Chat Tentatively Hallucinated in Extended Conversations with Users &
Bing blogs with iterations to safeguarding features \\
\hline
*UnitedHealth Accused of Deploying Allegedly Flawed AI to Deny Medical
Coverage & 
Victims raise with courts \\
\hline
ChatGPT Reportedly Produced False Court Case Law Presented by Legal
Counsel in Court & 
Discovered by courts \\
\bottomrule
    \end{tabular}
    \caption{Incidents with responses where the deployer and the developer are known but they are different organisations. Responses marked with a * where existing accountability mechanisms are employed in cases where vulnerable individuals were harmed.}
\end{table}

\newpage
\section{Incident 597 reports and response timeline}
\begin{figure}[h]
    \centering
    \includegraphics[width=\linewidth]{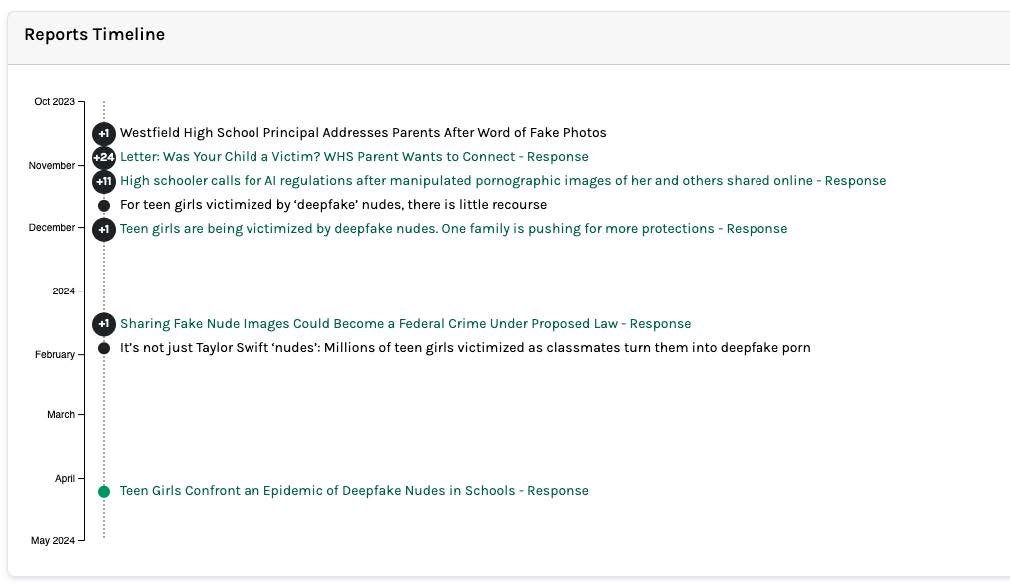}
    \caption{Incident 597 reports and response timeline. Copied from the AIID website}
\end{figure}
\label{apx:597}

\end{document}